\documentclass[twoside,twocolumn,9pt]{article}
\usepackage{extsizes}
\usepackage[super,sort&compress,comma]{natbib} 
\usepackage[version=3]{mhchem}
\usepackage[left=1.5cm, right=1.5cm, top=1.785cm, bottom=2.0cm]{geometry}
\usepackage{balance}
\usepackage{braket}
\usepackage{widetext}
\usepackage{times,mathptmx}
\usepackage{sectsty}
\usepackage{graphicx} 
\usepackage{lastpage}
\usepackage[format=plain,justification=raggedright,singlelinecheck=false,font={stretch=1.125,small,sf},labelfont=bf,labelsep=space]{caption}
\usepackage{float}
\usepackage{fancyhdr}
\usepackage{fnpos}
\usepackage[english]{babel}
\usepackage{array}
\usepackage{droidsans}
\usepackage{charter}
\usepackage[T1]{fontenc}
\usepackage[usenames,dvipsnames]{xcolor}
\usepackage{setspace}
\usepackage[compact]{titlesec}
\usepackage{color}
\usepackage{hyperref}
\usepackage{tikz}
\usetikzlibrary{shapes}

\usepackage{epstopdf}

\definecolor{cream}{RGB}{222,217,201}

\begin{document}

\pagestyle{fancy}
\thispagestyle{plain}
\fancypagestyle{plain}{

\renewcommand{\headrulewidth}{0pt}
}

\makeFNbottom
\makeatletter
\renewcommand\LARGE{\@setfontsize\LARGE{15pt}{17}}
\renewcommand\Large{\@setfontsize\Large{12pt}{14}}
\renewcommand\large{\@setfontsize\large{10pt}{12}}
\renewcommand\footnotesize{\@setfontsize\footnotesize{7pt}{10}}
\makeatother

\renewcommand{\thefootnote}{\fnsymbol{footnote}}
\renewcommand\footnoterule{\vspace*{1pt}%
\color{cream}\hrule width 3.5in height 0.4pt \color{black}\vspace*{5pt}} 
\setcounter{secnumdepth}{5}

\makeatletter 
\renewcommand\@biblabel[1]{#1}            
\renewcommand\@makefntext[1]%
{\noindent\makebox[0pt][r]{\@thefnmark\,}#1}
\makeatother 
\renewcommand{\figurename}{\small{Fig.}~}
\sectionfont{\sffamily\Large}
\subsectionfont{\normalsize}
\subsubsectionfont{\bf}
\setstretch{1.125} 
\setlength{\skip\footins}{0.8cm}
\setlength{\footnotesep}{0.25cm}
\setlength{\jot}{10pt}
\titlespacing*{\section}{0pt}{4pt}{4pt}
\titlespacing*{\subsection}{0pt}{15pt}{1pt}

\fancyfoot{}
\fancyfoot[RO]{\footnotesize{\sffamily{1--\pageref{LastPage} ~\textbar  \hspace{2pt}\thepage}}}
\fancyfoot[LE]{\footnotesize{\sffamily{\thepage~\textbar\hspace{3.45cm} 1--\pageref{LastPage}}}}
\fancyhead{}
\renewcommand{\headrulewidth}{0pt} 
\renewcommand{\footrulewidth}{0pt}
\setlength{\arrayrulewidth}{1pt}
\setlength{\columnsep}{6.5mm}
\setlength\bibsep{1pt}

\makeatletter 
\newlength{\figrulesep} 
\setlength{\figrulesep}{0.5\textfloatsep} 

\newcommand{\topfigrule}{\vspace*{-1pt}%
\noindent{\color{cream}\rule[-\figrulesep]{\columnwidth}{1.5pt}} }

\newcommand{\botfigrule}{\vspace*{-2pt}%
\noindent{\color{cream}\rule[\figrulesep]{\columnwidth}{1.5pt}} }

\newcommand{\dblfigrule}{\vspace*{-1pt}%
\noindent{\color{cream}\rule[-\figrulesep]{\textwidth}{1.5pt}} }

\makeatother

\twocolumn[
  \begin{@twocolumnfalse}
\vspace{3cm}
\sffamily
\begin{tabular}{m{4.5cm} p{13.5cm} }

 & \noindent\LARGE{\textbf{Dipolar Exchange Quantum Logic Gate with Polar Molecules}} \\
\vspace{0.3cm} & \vspace{0.3cm} \\

 & \noindent\large{Kang-Kuen Ni,\textit{$^{abc}$}$^{\ast}$ Till Rosenband,\textit{$^{b}$} and David D. Grimes\textit{$^{abc}$}} \\

 & \noindent\normalsize{We propose a two-qubit gate based on dipolar exchange interactions between individually addressable ultracold polar molecules in an array of optical dipole traps. Our proposal treats the full Hamiltonian of the $^1\Sigma^+$ molecule NaCs, utilizing a pair of nuclear spin states as storage qubits. A third rotationally excited state with rotation-hyperfine coupling enables switchable electric dipolar exchange interactions between two molecules to generate an iSWAP gate. All three states are insensitive to external magnetic and electric fields.  Impacts on gate fidelity due to coupling to other molecular states, imperfect ground-state cooling, blackbody radiation and vacuum spontaneous emission are small, leading to potential fidelity above $99.99~\%$ in a coherent quantum system that can be scaled by purely optical means.} \\

\end{tabular}

 \end{@twocolumnfalse} \vspace{0.6cm}

  ]

\renewcommand*\rmdefault{bch}\normalfont\upshape
\rmfamily
\section*{}
\vspace{-1cm}


\footnotetext{\textit{$^{a}$~Department of Chemistry and Chemical Biology, Harvard University, Cambridge, Massachusetts, 02138, USA.}}
\footnotetext{\textit{$^{b}$~Department of Physics, Harvard University, Cambridge, Massachusetts, 02138, USA.}}
\footnotetext{\textit{$^{c}$~Harvard-MIT Center for Ultracold Atoms, Cambridge, Massachusetts, 02138, USA.}}
\footnotetext{$^{\ast}$\textit{ E-mail: ni@chemistry.harvard.edu}}


\section{Introduction}\label{sec. 1}

Important progress has been made towards a laboratory quantum computer with state-of-the art demonstrations reaching a combination of
5 qubits and 98.3~\% CNOT gate fidelity~\cite{Linke2017, Leung2018}.  The criteria for quantum computation~\cite{DiVincenzo2000} have been identified as (1) a scalable system of qubits (2) initialization (3) coherence (4) universal set of qubit gates (5) measurement.  Items 2 through 5 have been demonstrated at sufficient fidelities~\cite{Brown2011, Ballance2016, Gaebler2016, Sheldon2016, Barends2014}, showing that computation with many qubits~\cite{Calderbank1996,Steane1996,Preskill1998} may be possible.  But the route toward scalability remains challenging.  Here, we focus on the problem of producing a high-fidelity two-qubit gate using optically trapped dipolar molecules, with the hope that this physical system allows for easier scalability.  Recent demonstrations of flexible optical tweezer arrays~\cite{Barredo2016, Endres2016} show a method by which many qubits could be rearranged to implement quantum algorithms.

Optically trapped, electrically dipolar neutral molecules have long been recognized as potential qubits~\cite{DeMille2002, Yelin2006,Zhu2013,Herrera2014,Karra2016} where the dipole-dipole interaction between two molecules mediates a two-qubit gate. However, most proposals rely on static or oscillating dressing electric fields to polarize the molecules, where the molecular Stark energy is much larger than the dipolar interaction.  This imposes stringent constraints on field stability. 

Here we describe concretely how the natural dipolar interaction between two molecules can produce an iSWAP gate, without the need for additional polarizing fields, thereby removing an important source of implementation complexity and qubit decoherence.
This iSWAP gate, together with single qubit rotations, forms a universal set of qubit gates~\cite{DiVincenzo1995, Schuch2003}.  We exploit the rich molecular internal structure and use NaCs as an example to find molecular qubits that are  expected to have long coherence (item 3).  The gate relies on two-qubit interactions that are switched on by driving one qubit state to a third state via a microwave transition. We find parameters that allow gates with high fidelity ($F>1-10^{-4}$) in 10~ms at a magnetic field of 1~Gauss when light shifts due to the optical trap are neglected. For an optical trap depth of 600 kHz, the same fidelity and duration can be reached for a 35 Gauss magnetic field.  The gate duration could be reduced by applying shaped pulses rather than the square pulses considered here.

\section{Exchange and the iSWAP gate}\label{seciSWAP}
It is a well known phenomenon that if two identical systems interact weakly, where one system has an energy excitation and the other does not, the excitation eventually transfers.  This effect can form the basis for a two-qubit gate~\cite{Loss1998,Mozyrsky2001,Schuch2003, Blais2004, Andre2006, Herrera2014} and has been discussed in the context of molecule-based quantum simulations of spin models~\cite{Barnett2006,Gorshkov2011a}.  The transfer of excitation via the dipole-dipole interaction has been demonstrated for ultracold KRb molecules~\cite{Yan2013,Wall2015}  and atoms with large magnetic dipoles~\cite{dePaz2013} in optical lattices.

\begin{figure*}
\centering
  \includegraphics[width=\linewidth]{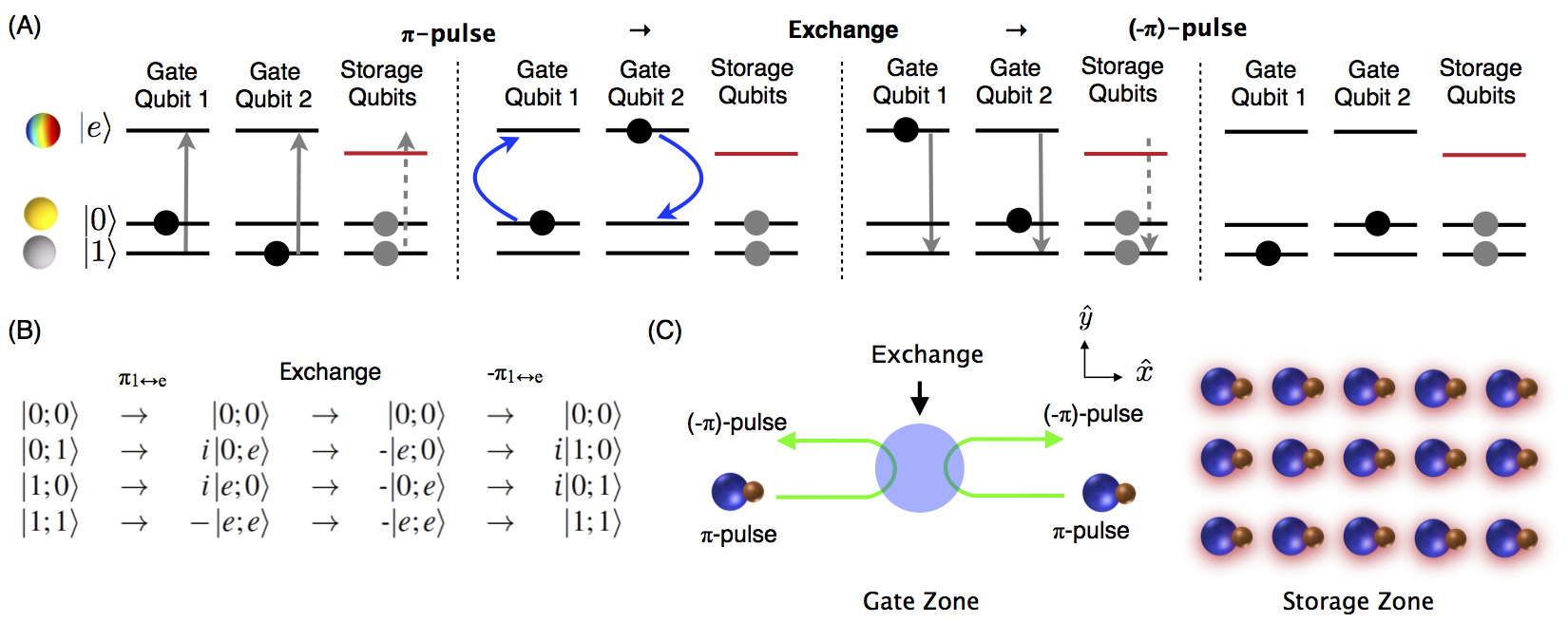}
\caption{ (A) iSWAP gate based on dipolar exchange between a pair of molecular states with opposite parity.  The colored sphere of the $\ket{e}$ state represents the wavefunction amplitude of the dipole direction for an $N=1,m_N=0$ state, where the quantization axis is horizontal. The states $\ket{0}$ and $\ket{1}$ are hyperfine sublevels of the rotational ground state $N=0$.  Superpositions of $\ket{e}$ and $\ket{0}$ or $\ket{1}$ produce an electric dipole moment that oscillates at a frequency corresponding to approximately twice the rotational constant of the molecule and couples to a nearby qubit. The four panels in (A) show the initial state $\ket{0;1}$ evolving through the gate to $i\ket{1;0}$, where horizontal arrows indicate the flow of time. State $\ket{e}$ of the storage qubits is light shifted out of resonance, for individual addressability. (B) includes other basis states and important details of quantum phases.  (C) Qubit array based on molecular hyperfine states.  Any pair of qubits can be moved from the storage zone to the gate zone in a flexible array of optical tweezers.  During the gate operation, a spatially uniform microwave pulse transfers population from state $\ket{1}$ to $\ket{e}$ in the gate qubits, so that the amplitudes of the $\ket{0;e}$ and $\ket{e;0}$ states are exchanged.  To achieve individual addressability with high spatial resolution, the light that shifts the storage qubits out of resonance (indicated by red shadows) can be produced in a similar way as the tweezer array.}
\label{figSetup}
\end{figure*} 

We rely on the natural dipole-dipole interaction between molecules to enable evolution of the type $\ket{0;e}\leftrightarrow\ket{e;0}$ where $\ket{0}$ is a sub-level of the rotational ground state, and $\ket{e}$ is a sub-level of the first rotational excited state (see Fig.~\ref{figSetup}A,B). In the two-particle state, the first position refers to the first molecule, and the second position refers to the second molecule.  This exchange interaction leaves the states $\ket{0;0}$ and $\ket{e;e}$ unchanged, and the overall unitary evolution in the basis $\ket{0;0},\ket{0;e},\ket{e;0},\ket{e;e}$ is~\cite{Wall2015}

\begin{equation}
\label{eqPropagator}
    \hat{U} = e^{-i\hat{H}t/\hbar} = \left(\begin{array}{c c c c}
1 & 0 & 0 & 0 \\
0 & \cos \Omega t & i\sin \Omega t & 0 \\
0 & i\sin \Omega t & \cos \Omega t & 0 \\
0 & 0 & 0 & 1 \\\end{array}\right),
\end{equation}
where $\Omega=D/r^3$ is the interaction Rabi rate, $r$ is the molecule-molecule distance, and the duration $t=\pi/(2 \Omega)$ produces the iSWAP gate.  The factor $D$ (see Eq.~\ref{eqHDD}) depends on the choice of molecule, separation direction, choice of states, magnetic field, and light shift.  In the $1$~Gauss example of Section~\ref{secGateNumbers}, two NaCs molecules have $t=4$~ms for $r=2.9~\mu$m separation along $\hat{x}$, which is also the magnetic field direction.  For the 35~Gauss example, which includes the effects of a 600 kHz deep optical trap, we use an interaction duration of $t=2$~ms with $r=2.5~\mu$m along $\hat{x}$ and the magnetic field direction is $\hat{z}$.  While the resolution of optical tweezers with beam waist below $1~\mu$m supports smaller separation and gates as fast as $t=50~\mu$s, two effects place additional constraints.  1. Off-resonant population leakage degrades the gate fidelity for short durations (Sec.~\ref{secLight}) and 2. smaller molecule separation ($r$) makes the gate more sensitive to motional excitation (Sec.~\ref{secMotion}).

Quantum computing requires two qubit states $\ket{0}$ and $\ket{1}$ that are coherent and couple minimally to the environment and other qubits.  In this proposal, we utilize two hyperfine sublevels of the rotational ground state of a molecule as states $\ket{0}$ and $\ket{1}$.  Long-lasting coherence of such states has recently been demonstrated in a gas of ultracold NaK molecules~\cite{Park2017}.  While the hyperfine levels offer coherence, they do not produce strong dipole-dipole interactions between nearby molecules.  To enable this interaction and produce an iSWAP gate, the $\ket{1}$ state in two molecules is temporarily transferred to the rotationally excited state $\ket{e}$ via a microwave $\pi$-pulse.  Then, after energy exchange, the $\ket{e}$ population is transferred back to $\ket{1}$.  The propagator in Eq.~\ref{eqPropagator} still applies, but now in the computational basis $\ket{0;0},\ket{0;1},\ket{1;0},\ket{1;1}$ as shown in Fig.~\ref{figSetup}(B).

The above sequence requires an excited state $\ket{e}$ that couples to two different hyperfine levels ($\ket{0}$ and $\ket{1}$) of the ground state via electric fields.  We rely on the hyperfine interaction term of the internal molecular Hamiltonian that couples molecular rotation to nuclear spin via the nuclear electric quadrupole moment~\cite{Zare1988,Aldegunde2017} to produce eigenstates that contain superpositions of different nuclear spin states.  This interaction requires a nuclear spin greater than 1/2.  Even though external electric fields do not change the nuclear spin directly, they can change the nuclear spins by driving transitions between states with different superpositions~\cite{Aldegunde2009, Ospelkaus2010a}.  
In this manuscript, we use $^{23}$Na$^{133}$Cs as an example because it has a large permanent electric dipole moment (4.6 Debye), and full quantum control of individually trapped molecules is being developed~\cite{Liu2017, Liu2018}.  A similar gate scheme that makes use of internal molecular couplings could also be applied to other ultracold polar molecules, including other bialkalis where a single internal quantum state can already be prepared~\cite{Ospelkaus2010a, Takekoshi2014, Will2016,Gregory2017, Guo2018} and molecules of $^2\Sigma$ electronic structure with spin-rotational coupling~\cite{Barry2014,Truppe2017,Anderegg2017,Anderegg2018}.

\begin{figure}[h]
\centering
\includegraphics[width=0.9\linewidth]{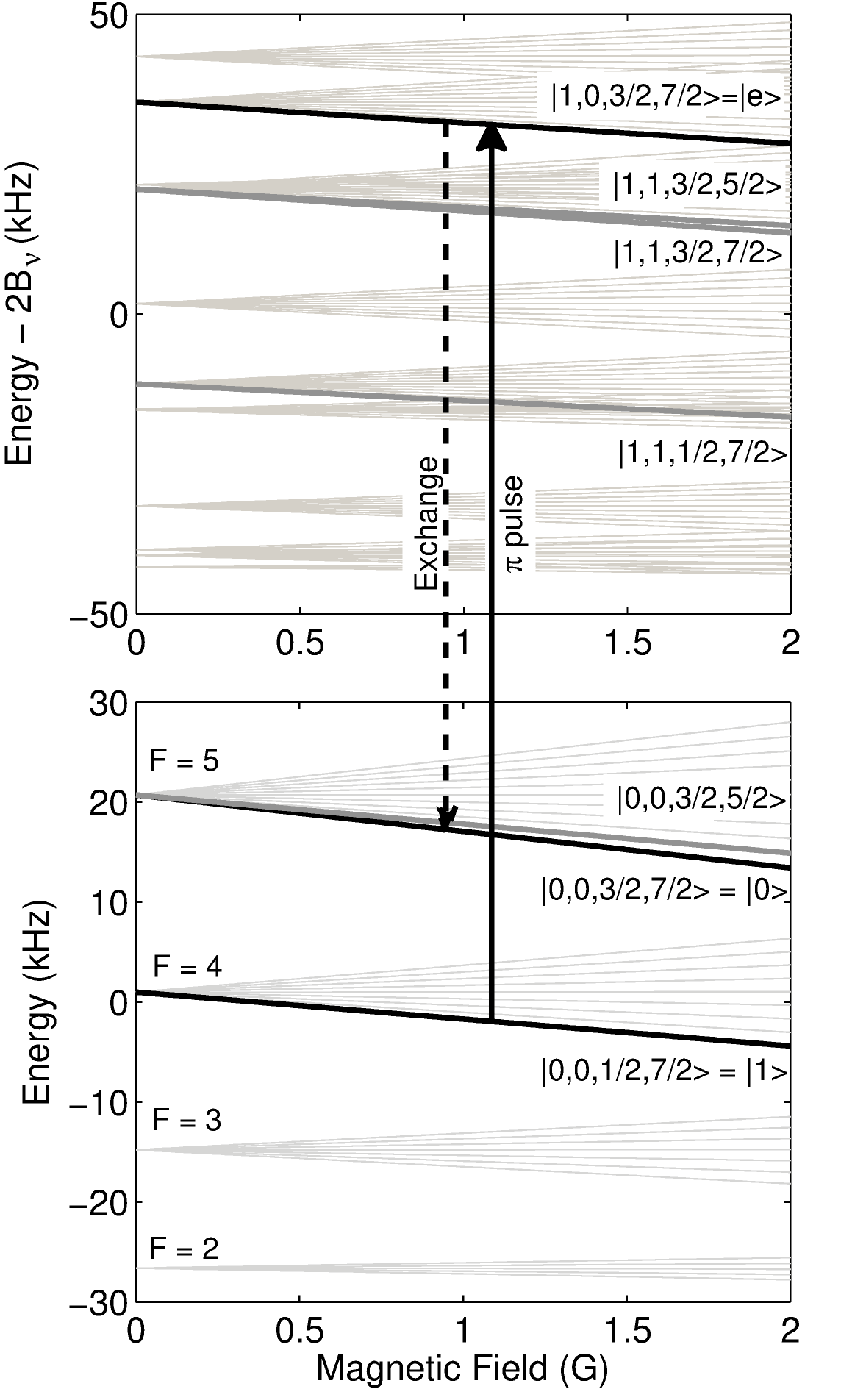}
\caption{Hyperfine and Zeeman energy levels as a function of magnetic field for the $N=1$ (top) and $N=0$ (bottom) states of $^{23}$Na$^{133}$Cs ($v=0$) in zero electric field. While a number of states are nearly degenerate with the states of interest $\ket{0},\ket{1}, and \ket{e}$, selection rules prevent them from participating in the interactions.  $B_v=1.7396$~GHz is the molecular rotation constant.}
\label{figLevels}
\end{figure}

\section{Qubit array}
For a quantum computer, a large number of molecular qubits could be held in an array of optical tweezers for storage (see Fig.~\ref{figSetup}C).  Two arbitrary qubits can be selected for gate operations by means of a configurable tweezer array~\cite{Barredo2016,Endres2016} and moved so they are separated from the other qubits and initially far from each other.  Off-resonant light is applied to the array of stored qubits to shift their $\ket{e}$ energy level so that the $\pi$-pulses have no effect. Individual addressability of qubits requires a spatial light pattern with high contrast ratio between stored and gate qubits, which can be generated by similar optics as the array of optical tweezers.  The light causes only very small differential shifts for the storage states $\ket{0}$ and $\ket{1}$.  A spatially uniform microwave $\pi$-pulse of well-defined polarization then transfers $\ket{1}$ to $\ket{e}$ for the two gate qubits. Then the qubits are moved near each other and back apart to allow for the interaction.  This movement naturally produces a temporally shaped interaction with gradually rising and falling strength to reduce off-resonant population leakage. The final $\pi$-pulse transfers the  $\ket{e}$ population to $\ket{1}$.  An enhanced gate sequence can include a central spin-echo pulse to mitigate slowly drifting energies (Sec.~\ref{secLight}).  More advanced dynamical-decoupling sequences can be applied to further reduce decoherence and the sensitivity to qubit motion (Sec.~\ref{secMotion}).

\section{Gate calculation}
\label{secCalc}
Because the molecule is in a $^1\Sigma^+$ state, the electronic spin and orbital angular momentum are zero, and do not enter the calculation.  We identify quantum states in the uncoupled basis by the quantum numbers $N,m_N,m_1$, and $m_2$, where $N$ is the angular momentum associated with molecular rotation and $m_N,m_1,m_2$ are projections of this angular momentum and the two nuclear spins onto the magnetic field axis. While the uncoupled basis is convenient for calculations, the basis states generally do not coincide with eigenstates of the molecular Hamiltonian.  Nevertheless, we label Hamiltonian eigenstates as $\ket{N,m_N,m_1,m_2}$, where we use the quantum numbers of the uncoupled basis state with maximum overlap. Although this labeling scheme could in principle assign the same quantum numbers to two different eigenstates, we have verified that this does not occur for the specific states discussed here.  When two-molecule states are described, they are written as $\ket{a;b}$ where $a$ and $b$ are the states of molecules 1 and 2 respectively.

We solve for the eigenstates associated with the molecular hyperfine Hamiltonian for NaCs~\cite{Aldegunde2017} at various magnetic fields and optical trap depths.  The energy levels are shown as a function of magnetic field in Fig.~\ref{figLevels} and trap depth in Fig.~\ref{figLightShift}.  The dipole-dipole interaction and electric-field-driven $\pi$-pulses both depend on the electric dipole moments of the molecules. The Hamiltonian associated with an externally applied electric field to molecule $j$ is 
\begin{equation}
\hat{H}_E=-\mathbf{\hat{d}_j \cdot E}
\label{eqHE}
\end{equation} where we use the interaction picture and rotating-wave approximation to remove the time-dependence of  oscillating fields.  The dipole-dipole interaction is 
\begin{equation}
\hat{H}_{DD}=\frac{1}{4 \pi \epsilon_0 r^3}\left[\mathbf{\hat{d}_1\cdot\hat{d}_2}-
3(\mathbf{\hat{d}_1\cdot\hat{e}_r})(\mathbf{\hat{d}_2\cdot\hat{e}_r})\right]
\label{eqHDD}
\end{equation}
where $\mathbf{\hat{d}_j}$ is the dipole moment operator of molecule $j$.  Here, $\mathbf{\hat{e}_r}$ is the unit vector along the separation direction for the two molecules and the dot products are evaluated as sums over the three spatial directions.  The dipole moment operators are determined by the
rotational part of the energy eigenstates and for a basis state $N,m_n$, this axis has an orientation whose quantum wave function is the spherical harmonic $Y^N_m$. From the eigenstates of the molecular Hamiltonian we calculate the matrices associated with the dipole operators $\hat{d}_x, \hat{d}_y, \hat{d}_z$ for the laboratory coordinate system~\cite{Wall2015}.  These matrix elements are diagonal in the quantum numbers $m_1$ and $m_2$ of the uncoupled basis and can be reduced to Wigner-3j symbols~\cite{Zare1988}.

For each step ($\pi$-pulse and exchange), we calculate unitary evolution according to the time-independent Hamiltonians in Eq.~\ref{eqHE} and \ref{eqHDD}.  The resonant couplings drive the desired gate behavior, while off-resonant couplings to other states cause population leakage and Stark shifts.  For a low magnetic field without light, the most important Hamiltonian terms are listed in Tables~\ref{tblHpi} and ~\ref{tblHX}.  The unitary transformation is applied to many test states in the computational basis to determine the minimum gate fidelity~\cite{NielsenChuang2000}.  The approximate population leakage can also be calculated more simply via perturbation theory (Appendix~\ref{secLeakage}).

\begin{table}
\centering
\begin{tabular}{c|c|r|r}
    From ($i$) & To ($j$) & $H_{ij}/\hbar$ [s$^{-1}$] & $H_{jj}/\hbar$ [s$^{-1}$]\\
    \hline \hline
$\ket{e}$ & $\ket{1}$ & 521.9 & 0 \\
$\ket{1}$ & $\ket{e}$ & 521.9 & 6.3 \\
$\ket{0}$ & $\ket{0}$ & - & 117851.0 \\
\hline
$\ket{1}$ & $\ket{1,1,1/2,7/2}$ & 1135.6 & -285144.0 \\
$\ket{1}$ & $\ket{1,1,3/2,5/2}$ & 2.3 & -82709.5 \\
$\ket{e}$ & $\ket{0,0,3/2,5/2}$ & 470.8 & 122454.0 \\
$\ket{0}$ & $\ket{1,1,3/2,7/2}$ & 1249.8 & -86537.2 \\
\end{tabular}
\caption{Non-zero coupling terms of the $\ket{1}\leftrightarrow\ket{e}$ $\pi$-pulse Hamiltonian (interaction picture) in the rotating-wave approximation when $\sigma_+$-polarized radiation at $3.48$~GHz is applied to a molecule in $1$~Gauss magnetic field.  The electric field amplitude is 0.03~V/m, such that the $\pi$-pulse duration is 3.01~ms (adjusted for maximum fidelity).  The radiation frequency has been adjusted by 6.3 radians/s to compensate for dynamic Stark shifts.  All terms connecting to $\ket{0}$, $\ket{1}$, or $\ket{e}$ are shown.}
\label{tblHpi}
\end{table}

\begin{table}
\centering
\begin{tabular}{c|c|r|r}
    From & To & $H_{ij}/\hbar$& $H_{jj}/\hbar$\\
    $(i)$ & $(j)$ & [s$^{-1}$] & [s$^{-1}$] \\
    \hline \hline
$\ket{e;0}$ & $\ket{0;e}$ & -390.7 & 0 \\
\hline
$\ket{e;0}$ & $\ket{0;1,1,3/2,5/2}$ & -195.6 & -82715.8 \\
$\ket{e;0}$ & $\ket{0,0,3/2,5/2;1,1,3/2,7/2}$ & 107.6 & -81940.9 \\
$\ket{e;0}$ & $\ket{0;1,1,1/2,7/2}$ & 180.0 & -285151.0 \\
$\ket{e;0}$ & $\ket{1;1,1,3/2,7/2}$ & 119.3 & -204395.0
\end{tabular}
\caption{Non-zero coupling terms for $\ket{e;0}$ of the exchange Hamiltonian when two molecules are separated by $2.9~\mu$m along the $1$~Gauss magnetic field. Propagation of the Hamiltonian approximates the exchange evolution (Eq. 1) when applied for a duration 4.02~ms (adjusted for maximum fidelity). 
The terms $\ket{0;0}$ and $\ket{e;e}$ do not couple to other states.}
\label{tblHX}
\end{table}

\section{Gate speed and fidelity}
\label{secGateNumbers}
In addition to the desired evolution in Eq.~\ref{eqPropagator}, the iSWAP gate sequence causes off-resonant coupling to other molecular states.  This causes a trade-off between gate speed and fidelity.  Although two NaCs molecules could be brought to a separation below $1~\mu$m, with an exchange duration of $50~\mu$s, such an interaction would limit the gate fidelity to $99.6$~\%.  At low magnetic field with NaCs, we find that the interactions couple off-resonantly to other states with $\delta>10$~kHz detuning both for the $\pi$-pulses and the exchange.  For the square pulses considered here, time-energy uncertainty causes off-resonant population transfer out of the computational basis states and a gate error of order $(\tau \delta)^{-2}$.  Off-resonant coupling to other states also causes dynamic Stark shifts which result in small reproducible phase and frequency shifts. We expect that these can be corrected for without loss of fidelity.

\begin{figure*}
\centering
  \includegraphics[width=.9\linewidth]{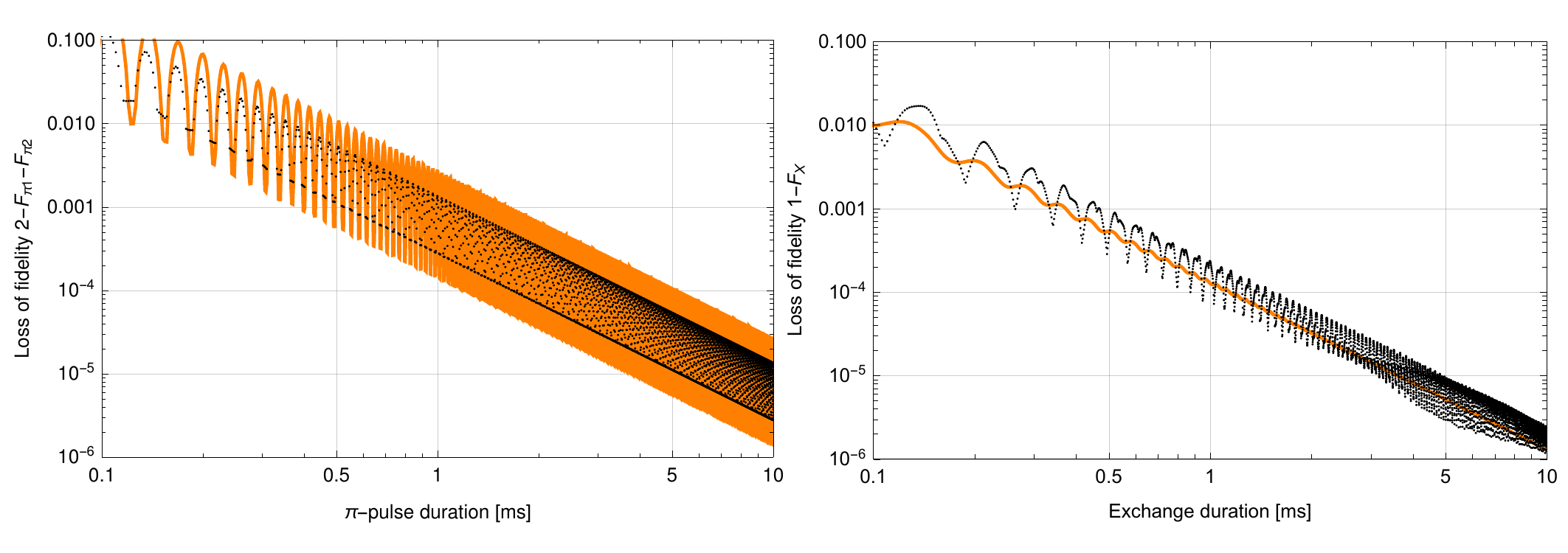}
\caption{Loss of fidelity as a function of duration for the two $\pi$-pulses ($2-F_{\pi1}-F_{\pi2}$) and exchange ($1-F_X$).  In both cases, the interaction strength (drive strength or molecule-molecule separation) is adjusted so the operation completes in the nominal duration.  The ripples are caused by the square pulse shape whose sinc-function power-spectrum-minima  move across off-resonant transitions as the duration is varied.  The scaling behavior for the maximum fidelity points is $2-F_{\pi1}-F_{\pi2}=2.8\times10^{-4}t^{-2}$ and $1-F_{x}=8.5\times10^{-5}t^{-2}$ where $t$ is given in ms. Details of the fidelity calculation are given in Section~\ref{secCalc}.  The solid lines represent the perturbation theory results while the dotted lines correspond to the full Hamiltonian.}
\label{figFidelity}
\end{figure*}

We have calculated the unitary time evolution for all states that could couple via microwave pulses ($\hat{H}_E$) and dipole-dipole interactions ($\hat{H}_{DD}$) to the states in Fig~\ref{figSetup}(B). For low magnetic field (1 Gauss), details of the Hamiltonian terms are given is Tables~\ref{tblHpi} and \ref{tblHX}, where we have chosen the states $\ket{0}=\ket{0,0,3/2,7/2}$, $\ket{1}=\ket{0,0,1/2,7/2}$ and $\ket{e}=\ket{1,0,3/2,7/2}$. 
  
We find that a gate with 10~ms duration has a fidelity of $F = 1 - 3.6\times 10^{-5}$.  The exchange part of the example gate holds the two molecules at a distance of $2.9~\mu$m for a duration $\tau_x=4.02$~ms.  The $\pi$-pulses of duration $3.01$~ms utilize a $\sigma_+$ polarized microwave electric field at a frequency of $3.48$~GHz and amplitude $0.03$~V/m.  The $\pi$-pulse and exchange fidelities due to population leakage are shown separately in Figure~\ref{figFidelity}.  Population leakage out of the computational basis must be corrected for to support long gate sequences~\cite{Fowler2013}.  Off-resonant coupling can likely be reduced by use of shaped pulses rather than square ones, to reach the same fidelity in a shorter duration~\cite{Motzoi2009}.  At a higher magnetic field of 35~Gauss with $600$~kHz trap depth, the best states ($F=1-6\times10^{-5}$ at 9.4 ms) are $\ket{1}=\ket{0,0,3/2,5/2}$ and $\ket{e}=\ket{1,-1,3/2,7/2}$, with $\ket{0}$ as above.

\section{Intensity fluctuations and light shifts}
\label{secLight}
\begin{figure}
  \includegraphics[width=1.0\linewidth]{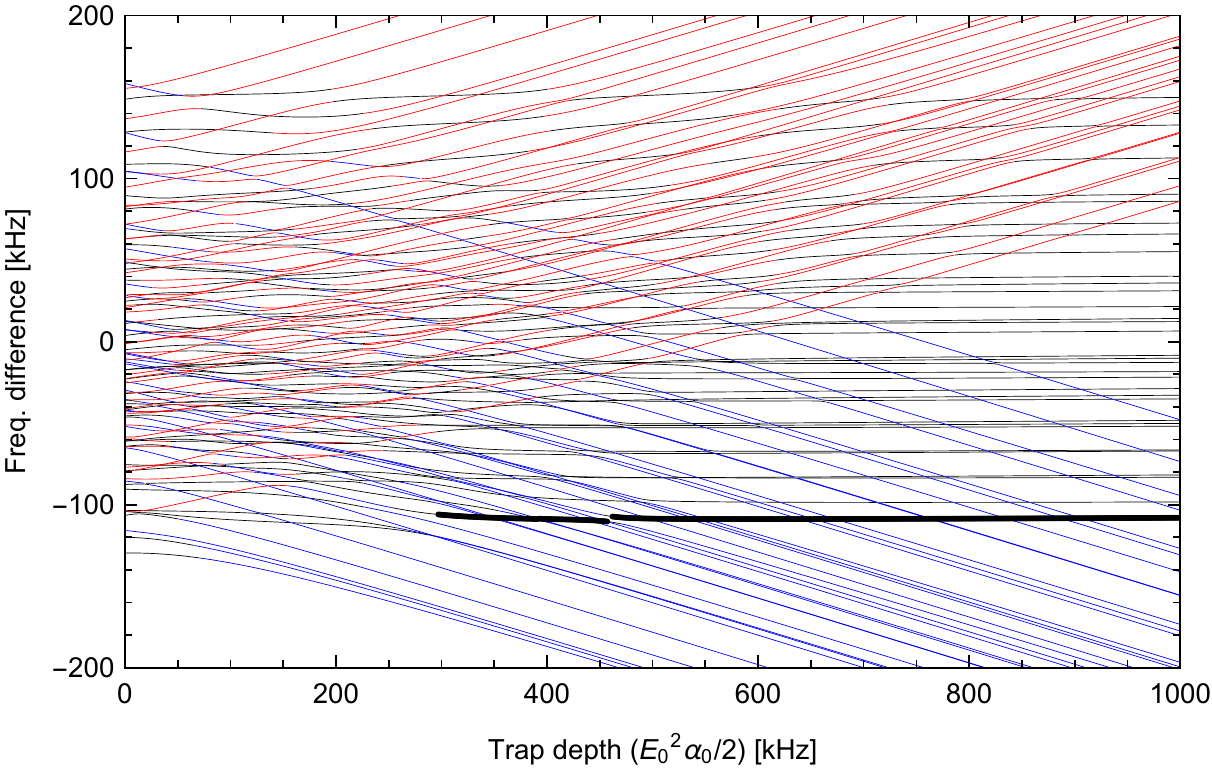}
\caption{Differential light shifts of excited rotational states ($N=1$) with respect to the ground state, as a function of trap depth.  The large offset of about 3.48~GHz has been subtracted.  States with predominantly $(\ket{m_N=-1}-\ket{m_N=1})/\sqrt{2}$ character with dipole moment along $\hat{x}$ are shown in black.  States with predominantly $(\ket{m_N=-1}+\ket{m_N=1})/\sqrt{2}$ character with dipole moment along $\hat{y}$ are blue.  States with predominantly $\ket{m_N=0}$ character with dipole moment along $\hat{z}$ are red. The thick line shows the state $\ket{e}=\ket{1,-1,3/2,7/2}$.  The polarization vector is $\hat{\epsilon}=\hat{x}\cos{\gamma}+i \hat{y}\sin{\gamma}$ with ellipticity $\gamma=35.6091^\circ$, adjusted to null the slope of $\ket{e}$ at 600~kHz.  A magnetic field of 35~Gauss lies along $\hat{z}$.}
\label{figLightShift}
\end{figure}

\begin{figure}
\centering
\includegraphics[width=.95\linewidth]{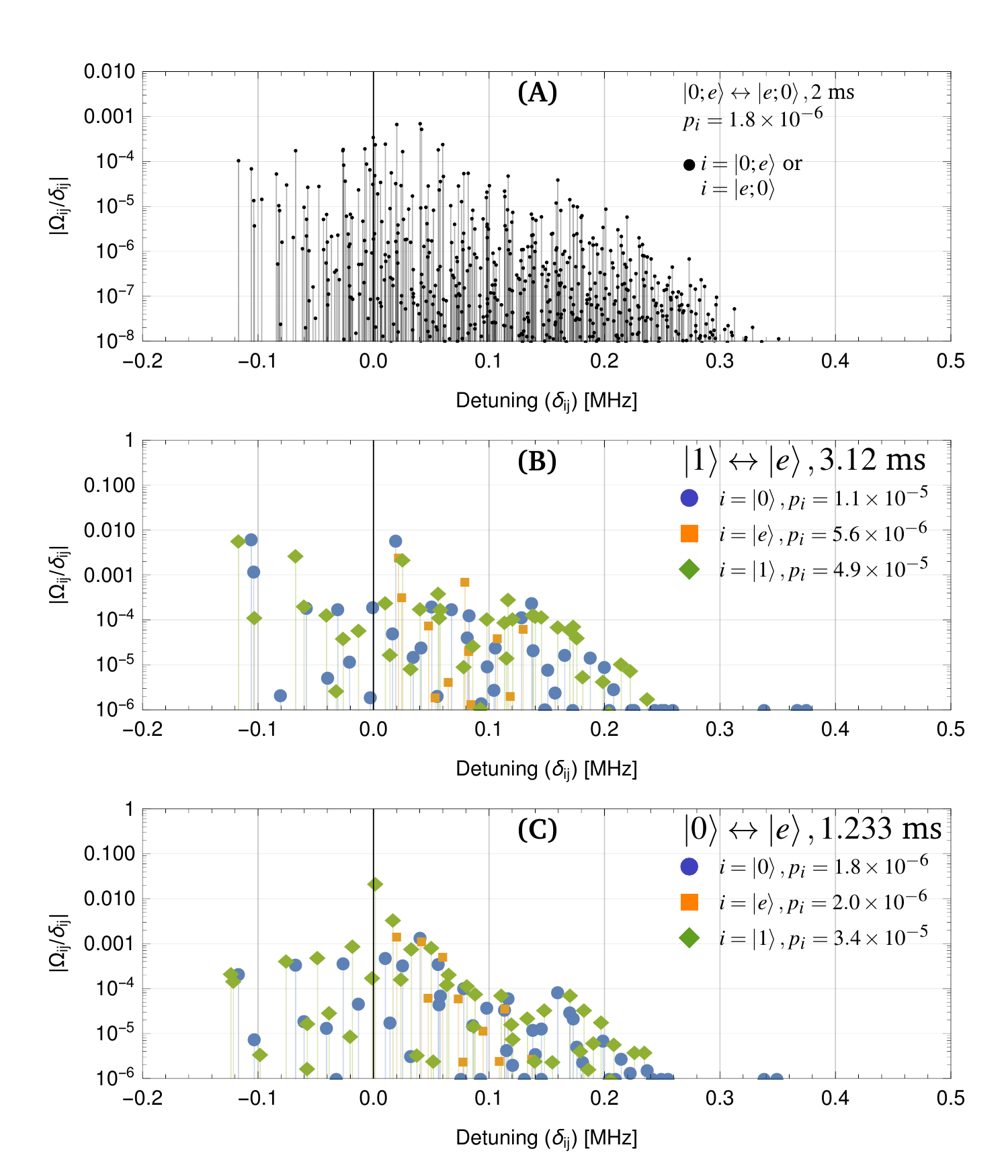}
\caption{Off-resonant coupling terms $|\Omega_{ij}/\delta_{ij}|$ that cause population leakage during the gate steps.  Here, $\Omega_{ij}$ is the coupling Rabi rate between source state $i$ and leakage state $j$, and $\delta_{ij}$ is the frequency difference in radians/s.  For each step, the probability of population leakage $p_i$ is shown (see Appendix~\ref{secLeakage}), where the summation is over all coupled states $j$.  $p_i$ can be used to estimate off-resonant population leakage without calculating the full unitary time evolution.  {\bf (A)} Exchange interaction with separation of $2.5~\mu$m along $\hat{x}$ for a 2 ms exchange duration. The fidelity, calculated from unitary time evolution is $F=1-2.0\times10^{-6}$. {\bf (B)} $\ket{1}\leftrightarrow\ket{e}$ $\pi$-pulse with electric field amplitude $0.0544$~V/m along $\hat{z}$ for a 3.12~ms pulse duration.  The fidelity for a pair of $\pi$-pulses, calculated from the unitary time evolution is $F=1-5.4\times10^{-5}$.  {\bf (C)} $\ket{0}\leftrightarrow\ket{e}$ $\pi$-pulse with electric field amplitude 0.0157~V/m along $\hat{x}$ for a 1.233 ms pulse duration.  The state $\ket{1}$ can be neglected if its population has already been transferred to $\ket{e}$.}
\label{figOffResonant}
\end{figure}

If the molecule is optically trapped, light shifts significantly perturb the energy eigenstates and transition frequencies.  To reach the wavefunction spread described in Section~\ref{secMotion}, we assume a $600~$kHz trap depth (12.9~kW/cm$^2$ intensity) with elliptical polarization~\cite{Rosenband2018}.  We also assume a trapping laser wavelength of 1030~nm.  The resulting energy levels are shown in Figure~\ref{figLightShift} 
for a magnetic field of 35~Gauss.  The higher magnetic field was chosen, because the Zeeman splitting reduces off-resonant couplings during the $\ket{1}\leftrightarrow\ket{e}$ $\pi$-pulses that are induced by the light shift Hamiltonian.  Figure~\ref{figOffResonant} shows the magnitude of off-resonant coupling  terms to other molecular states. For interactions that turn on and off sharply, the loss of fidelity due to off-resonant population leakage is given by a sum of terms involving off-resonant coupling strengths and detunings (see Appendix~\ref{secLeakage}).  Overall, we find that the exchange interaction has reduced leakage compared to the 1 Gauss case, even for a faster interaction.  The $\ket{1}\leftrightarrow\ket{e}$ $\pi$-pulses cause slightly higher leakage, which may be reduced by shaped or DRAG pulse techniques~\cite{Motzoi2009}.

For an optical trap with elliptical polarization, the quadratic sensitivity to intensity is calculated to be $\Delta f=(\Delta I/I)^2\times12.6$~kHz, leading to gate errors of $5\times 10^{-6}$ for relative intensity fluctuations of $\Delta I / I = 2\times 10^{-3}$.  A more significant effect is polarization drift.  For quartz waveplates with a temperature sensitivity of retardance of $10^{-4}$/C, one may expect polarization ellipticity fluctuations of $10^{-5}$ and light shift fluctuations of $4\times10^{-6}$ (2.5 Hz).  Because these fluctuations are slow thermal effects, spin echo pulses can reduce their impact.  If the quadratic light shift changes slowly, e.g. due to recoil heating, the effect is also mitigated.  One possible implementation is shown in Table~\ref{tblSpinEcho} where a $\ket{0}\leftrightarrow\ket{e}$ $\pi$-pulse is inserted in the middle of the gate sequence.  This leads to first-order insensitivity of the gate error with respect to constant light shift errors.  The result is equivalent to an iSWAP gate between the qubits, followed by inversion of the individual qubits.  Note that this spin-echo example was chosen for simplicity, and other dynamical decoupling techniques to compensate slow drifts could also be applied.  The $\ket{0}\leftrightarrow\ket{e}$ $\pi$-pulses, which are used for the spin-echo have little off-resonant coupling, even for a 1~ms duration (see Fig.~\ref{figOffResonant}).

\begin{table*}
\centering
\begin{tabular}{rcrcrcrcrcr}
    & $\pi_{1\leftrightarrow e}$ & & $\frac{1}{2}$Ex. & & $\pi_{0\leftrightarrow e}$ & & $\frac{1}{2}$Ex. & & $(-\pi_{1\leftrightarrow e})$ & \\
    $\ket{0;0}$ & $\rightarrow$ & $\ket{0;0}$ & $\rightarrow$ & $\ket{0;0}$ & $\rightarrow$ & $-\ket{e;e}$ &
    $\rightarrow$ & $-\ket{e;e}$ & $\rightarrow$ & $\ket{1;1}$ \\
    $\ket{0;1}$ & $\rightarrow$ & $i\ket{0;e}$ & $\rightarrow$ & 
    $\frac{i}{\sqrt{2}}\ket{0;e}-\frac{1}{\sqrt{2}}\ket{e;0}$ & $\rightarrow$ & 
    $\frac{1}{\sqrt{2}}\ket{0;e}-\frac{i}{\sqrt{2}}\ket{e;0}$ & $\rightarrow$ & 
    $\ket{0;e}$ & $\rightarrow$ & $-i\ket{0;1}$ \\
    $\ket{1;0}$ & $\rightarrow$ & $i\ket{e;0}$ & $\rightarrow$ & 
    $\frac{-1}{\sqrt{2}}\ket{0;e}+\frac{i}{\sqrt{2}}\ket{e;0}$ & $\rightarrow$ & 
    $\frac{-i}{\sqrt{2}}\ket{0;e}+\frac{1}{\sqrt{2}}\ket{e;0}$ & $\rightarrow$ & 
    $\ket{e;0}$ & $\rightarrow$ & $-i\ket{1;0}$ \\
    $\ket{1;1}$ & $\rightarrow$ & $-\ket{e;e}$ & $\rightarrow$ & $-\ket{e;e}$ & $\rightarrow$ & $\ket{0;0}$ &
    $\rightarrow$ & $\ket{0;0}$ & $\rightarrow$ & $\ket{0;0}$\\
\end{tabular}
\caption{Evolution of the computational basis states through the gate, which includes a central spin-echo pulse to cancel the phase evolution from slowly varying energy shifts between $\ket{0}$ and $\ket{e}$, such as light shifts.  The exchange interaction is split into two parts, where each has one half the duration required for full exchange.}
\label{tblSpinEcho}
\end{table*}

The differential light shift between two hyperfine ground states for NaK was measured~\cite{Park2017} as $5\times10^{-5}$~Hz/(W/cm$^2$).  With the assumption that the states $\ket{0}$ and $\ket{1}$ in NaCs have similar differential shifts, a beam with $10$~kW/cm$^2$ intensity and 1064~nm wavelength can Stark shift the $\ket{e}$ states of the stored qubits by $200$~kHz while shifting their $\ket{0}$ and $\ket{1}$ states by only $0.6$~Hz differentially.  In this case, the polarization is adjusted for maximum differential Stark shift of $\ket{e}$ with respect to $\ket{0}$ and a spatially patterned beam can be generated in the same way as the tweezer array. If the Stark shifting beam has a relative intensity stability of $2\times 10^{-3}$, the loss of fidelity is $1-F\approx 7\times 10^{-10}$ for a single stored qubit. 
The large ratio between $\ket{0}-\ket{e}$ and $\ket{0}-\ket{1}$ sensitivities makes it possible to individually address certain molecules by light-shifting the $\ket{e}$ state of the other molecules.

\section{Effects of molecule motion}\label{secMotion}
In the ideal gate, both molecules are in the motional ground state of their optical tweezer, and the exchange interaction strength is always the same.  However, ground-state cooling is imperfect and the molecules gain kinetic energy due to photon recoil.  Given the imaginary polarizability of NaCs Im$[\alpha]\approx 10^{-4}$ (atomic units)~\cite{Vexiau2017}, the total scattering rate for a 600~kHz deep trap is approximately $0.8$ Hz with a heating rate of $0.44$~quanta/s axially and $0.023$~quanta/s for each radial direction.  Therefore, it is desirable for the gate fidelity to exceed the ground-state occupation fidelity.  Here we examine the effect of motional excitations where the spatial coupling constant of Eq.~\ref{eqHDD} is modified.  For the proposed situation where the molecular dipoles lie along $\hat{x}$, this coupling scales as $(1-3\cos^2{\theta})/r^3$ where $\theta$ is the angle between the separation axis and $\hat{x}$, and $r$ is the separation distance (2.5~$\mu$m here).  For $\theta\approx0$, a series expansion in terms of spatial coordinates yields a change in the interaction Rabi rate $\Omega$ per motional quantum of
\begin{equation}
\label{eqMotionZ}
\Delta\braket{m|\Omega|m} / \Delta n_{j,k} \approx f_k\Omega_0 (s_k/r)^2
\end{equation}
where $\ket{m}=\ket{n_{1,x},n_{1,y},n_{1,z};n_{2,x},n_{2,y},n_{2,z}}$ is the state of two-molecule motion, $\Omega_0=D/r^3$ is the interaction Rabi rate without motion of Eq.~\ref{eqPropagator}, $n_{j,k}$ is the motional excitation number for molecule $j$ in direction $k\in\{x,y,z\}$, $s_k=\sqrt{\hbar/(2m \omega_k)}$ are the zero-point wave function spreads, $f_x=12$, $f_{y,z}=-6$, and the approximation includes terms up to second order in position.  The above trap depth corresponds to motional frequencies of $\omega_{x,y}=2\pi\cdot 12.4$ kHz and $\omega_z=2\pi\cdot 2.8$ kHz for NaCs, where a Gaussian beam radius of $1~\mu$m is assumed.  Because the loss of fidelity is proportional to the square of the Rabi rate deviation from the mean, $1-F$ can calculated from the variance of the Rabi rate, which can be expressed in terms of mean motional quantum numbers $\bar{n}_{j,k}$ for a thermal distribution of motional states:
\begin{equation}
\label{eqMotionF}
1-F  \approx \frac{\pi^2}{8} \sum_{j,k} f_k^2 (\bar{n}_{j,k}^2+\bar{n}_{j,k}) (s_k/r)^4.
\end{equation}
If the molecules are near the motional ground state with excitations dominated by imperfect cooling of Na~\cite{Yu2018}, the mean excitation numbers of each molecule along $x, y$, and $z$ are .019, .024, and .024 respectively.  Then motional effects cause a fidelity loss of $9\times10^{-6}$ for the above trap strength.  The effect of motion on the two-qubit gate can be reduced by implementing a BB1 sequence~\cite{Wimperis1994,Jones2003}.  Because the sign of $f_z$ is opposite $f_x$, it is possible to choose a separation direction that makes the interaction Rabi rate first-order insensitive to $n_{j,z}$.  This occurs for $\hat{r}=\hat{x}\cos{\theta}+\hat{z}\sin{\theta}$ with $\theta=(1/4)\cos^{-1}{(3/35)}$, and the interaction strength is reduced by 20~\%.  An optical lattice in the weakly confining direction would also reduce the effects of axial motion.  Note that the term $\bar{n}_{j,k}^2 s_k^4$ in Eq.~\ref{eqMotionF}, which limits fidelity for $\bar{n}_{j,k}>1$, is independent of trap intensity when $\bar{n}_{j,k}$ is dominated by recoil heating.  

The motional state also affects the average light shift, because the tweezer intensity drops  away from the trap center.  For a Gaussian beam trap with thermally excited molecules whose mean quantum numbers are $\bar{n}_k$, we find the relative intensity variance $var(I/I_0)\approx\sum_{k} g_k^2 (\bar{n}_{k}^2+\bar{n}_{k})$ where $g_k=-4(s_k/w)^2$ for the radial directions, $g_z=-2(s_z \lambda/(\pi w^2))^2$ and $w$ is the beam radius.  The associated standard deviation in relative intensity is $2\times 10^{-3}$ for the residual motion after ground-state cooling described above.  This corresponds to the intensity stability assumed in Section~\ref{secLight}.

\section{Other effects}
Effects from blackbody radiation and spontaneous emission are small for the system we have outlined above. The dominant effect at room temperature is due to blackbody radiation and for NaCs the vibrational transition absorption rate is $1.7 \times 10^{-3}$ s$^{-1}$ while the rotational transition rate is negligible~\cite{Vanhaecke2007}.  The spontaneous emission rate from the $N = 1$ rotationally excited state is of order $10^{-8}$ s$^{-1}$. 

The rate of decoherence due to scattering of optical tweezer photons has not been calculated.  However, where long coherence between hyperfine ground states of NaK was observed, the decoherence was attributed to spatial intensity variation in the optical trap rather than scattering~\cite{Park2017}.  A calculation of scattering rates should distinguish between internal-state-preserving (Rayleigh) and internal-state-changing (Raman) scattering, as the rate of Raman scattering can be several orders of magnitude lower than Rayleigh scattering~\cite{Cline1994,Uys2010}.

Magnetic field fluctuations cause dephasing of quantum states if their energies have an unequal slope with respect to field changes.  For both the 1 Gauss and 35 Gauss examples, we find that the relative sensitivities of the $\ket{0}$, $\ket{1}$, and $\ket{e}$ states are below 1~kHz/Gauss.  While this field sensitivity is small, the associated loss of fidelity grows with a factor $N^2$ if maximally entangled states such as $(\ket{0\cdots00}+\ket{1\cdots11})/\sqrt{2}$ with $N$ qubits are stored in the qubit array.  The same scaling applies to light that differentially shifts the phase of $\ket{0}$ and $\ket{1}$ (see Section~\ref{secLight}).  Such common-mode dephasing errors can be reduced by use of dynamical decoupling or decoherence-free subspaces~\cite{Zanardi1997}.

\section{Single qubit rotations}
Single qubit $X$ and $Y$ rotations are less complex than two-qubit gates and can be accomplished by individually addressing only one molecule and using a simplified sequence without dipole-dipole exchange.  It may be advantageous to first perform a $\pi$-pulse from $\ket{0}$ to $\ket{e}$, then a rotation between $\ket{1}$ and $\ket{e}$ and finally a $\pi$-pulse from $\ket{e}$ to $\ket{0}$ as shown in Table~\ref{tblEvolutionSingle}.  The $\ket{0} \leftrightarrow \ket{e}$ $\pi$-pulses can have less off-resonant coupling than the previously discussed $\ket{1} \leftrightarrow \ket{e}$ pulses (see Fig.~\ref{figOffResonant}), and high fidelity can be achieved for a shorter gate duration than in the two-qubit case.  Rotation about the $Z$ axis can be accomplished by a $\ket{0} \leftrightarrow \ket{e}$ $\pi$-pulse with one oscillator phase, followed by a second $\ket{0} \leftrightarrow \ket{e}$ $\pi$-pulse with a different oscillator phase.

\begin{table}
\centering
\begin{tabular}{rcrcrcr}
    & $\pi_{0\leftrightarrow e}$ & & Rot. & & $-\pi_{0\leftrightarrow e}$ & \\
    $\ket{1}$ & $\rightarrow$ & $\ket{1}$ & $\rightarrow$ & $\alpha\ket{1} - \beta\ket{e}$ & $\rightarrow$ & $\alpha\ket{1}+\beta\ket{0}$ \\
    $\ket{0}$ & $\rightarrow$ & $-\ket{e}$ & $\rightarrow$ & $\gamma\ket{1} - \delta\ket{e}$ & $\rightarrow$ & $\gamma\ket{1}+\delta\ket{0}$ 
    \end{tabular}
\caption{Single qubit rotation sequence where the $\pi$-pulses are performed as $Y$-rotations, and the ``Rotation" step acts on the states $\ket{1}$ and $\ket{e}$. } 
\label{tblEvolutionSingle}
\end{table}

\section{Conclusion}

We have described a room-temperature scheme for quantum computing based on iSWAP gates performed by dipolar molecules, individually trapped in optical tweezers.  Calculations indicate a potential gate fidelity above 0.9999 with low decoherence, although the rate of Raman scattering of optical tweezer photons remains to be determined. A modest magnetic field is used, and electric fields or field gradients are not needed.  Scaling to many qubits  would require an equal number of optical dipole traps in a movable pattern.  For this gate to be realized, individual neutral ground-state molecules must still be produced, and improved state-measurement is needed.  We expect that long gate sequences will require the mitigation of population leakage and recoil heating, e.g. by periodically teleporting the state of used molecules onto new ones~\cite{Fowler2013}.  While this proposal utilizes two hyperfine ground levels of NaCs, thirty other ground levels exist, which might allow each molecule to contain several qubits.

Although the proposed iSWAP gate is slower than quantum gates in other systems, decoherence effects can be small due to the fact that the qubit states are isolated from the environment by the symmetry of $^1\Sigma^+$ states.  The longer gate duration also reduces the noise bandwidth of actively stabilized parameters.  For the foreseeable future, experimental quantum computing will aim to increase the number of available qubits and gate fidelity, and molecules have the potential to advance both goals.

\bibliographystyle{rsc} 
\providecommand*{\mcitethebibliography}{\thebibliography}
\csname @ifundefined\endcsname{endmcitethebibliography}
{\let\endmcitethebibliography\endthebibliography}{}

\appendix
\section{Off-resonant coupling}
\label{secLeakage}
We wish to calculate the population leakage during gate interactions.  In this case, there are two ``main" states (here $\ket{a}$ and $\ket{b}$) and leakage states $\ket{j}$ to which the main states connect with large detuning and/or weak coupling.  For simplicity, we assume that the Hamiltonian associated with the interaction turns on and off instantaneously to produce time-independent coupling.  In a real implementation, it will be important to ramp the interactions up and down smoothly to minimize off-resonant coupling.

One approach for calculating leakage is to generate the full Hamiltonian matrix and diagonalize it to compute the unitary time evolution.  While straightforward, this is computationally expensive when there are many leakage states.  Complementary to the full calculation of unitary time evolution, we use first-order perturbation theory below to calculate the population leakage as a simple sum.  A comparison of the techniques can be seen in Fig.~\ref{figFidelity}.  Note that this does not account for coherent population buildup, which may develop if several interactions are combined without phase randomization.

During interactions between degenerate primary states $\ket{a}$ and $\ket{b}$ with coupled leakage states $\ket{j}$, the Hamiltonian can be written in terms of ``desired" coupling $\hat{H}_0$ and ``leakage" couplings $\hat{H}'$ as
\begin{eqnarray}
\hat{H} &=& \hat{H}_0+\hat{H}' \\
\hat{H}_0/\hbar &=& \Omega \ket{a}\bra{b} + \Omega\ket{b}\bra{a} + \sum_j \delta_j\ket{j}\bra{j}\\
\hat{H}'/\hbar &=& \sum_{i\in{\{a,b\}},j} (\Omega_{ij}\ket{i}\bra{j}+\Omega_{ij}\ket{j}\bra{i}).
\end{eqnarray} 
Here, the basis state phases have been chosen to make the coefficients $\Omega$ and $\Omega_{ij}$ real and positive.
The base Hamiltonian $\hat{H}_0$ has eigenvectors $\ket{+},\ket{-},\ket{j}$ with eigenvalues $\hbar\Omega,-\hbar\Omega,\hbar\delta_j$ and
\begin{eqnarray}
\ket{+} &=& (\ket{a} + \ket{b})/\sqrt{2} \\
\ket{-} &=& (\ket{a} - \ket{b})/\sqrt{2}.
\end{eqnarray}
According to first-order perturbation theory, the perturbed eigenstates (assuming that different $\ket{j}$ states don't couple to one another) are
\begin{eqnarray}
\ket{+'} &=& \ket{+} + \sum_{j} \alpha_j \ket{j} \\
\ket{-'} &=& \ket{-} - \sum_{j} \beta_j \ket{j} \\
\ket{j'} &=& \ket{j} - \alpha_j\ket{+} + 
                       \beta_j\ket{-}.
\end{eqnarray}
where $\alpha_j = \frac{\Omega_{aj}+\Omega_{bj}}{\sqrt{2}(\Omega-\delta_j)}$ and $\beta_j=\frac{\Omega_{aj}-\Omega_{bj}}{\sqrt{2}(\Omega+\delta_j)}$. These states approximately diagonalize the time-evolution operator and can be used to estimate the transitions from the initial state $\ket{a}$ into states $\ket{j}$ due to unitary time evolution $\hat{U}(t)=e^{-i \hat{H} t /\hbar}$ :
\begin{eqnarray}
\label{eqSumPerturbed}
\braket{j|\hat{U}(t)|a}&\approx&\sum_{k'}\braket{j|k'}\braket{k'|a}e^{-i t \braket{k'|\hat{H}/\hbar|k'}} \\
\label{eqLeakageGeneral}
 &\approx& \frac{1}{\sqrt{2}}e^{-it\delta_j}\left(\beta_j-\alpha_j\right) 
     + \frac{1}{\sqrt{2}}e^{-it\Omega} \alpha_j - \frac{1}{\sqrt{2}}e^{it\Omega} \beta_j
\end{eqnarray}
Where the Eq.~\ref{eqSumPerturbed} sum is over all perturbed states.  Simplification from Eq.~\ref{eqSumPerturbed} to Eq.~\ref{eqLeakageGeneral} utilizes unperturbed energies in the exponential terms.  In our case for any $j$, due to selection rules, only one of $\Omega_{aj}$ and $\Omega_{bj}$ is ever non-zero.  We call the non-zero value $\Omega_{j}$.  If one makes the assumption $t=\pi/(2\Omega)$ ($\pi$-pulse), the population leakage is 
\begin{equation}
\left|\braket{j|\hat{U}(t)|a}\right|^2 \approx \frac{|\Omega_{j}|^2 \left(\delta_j ^2-2 \delta_j  \Omega  \sin{\left(\delta_jt\right)}+\Omega ^2\right)}{\left(\delta_j ^2-\Omega ^2\right)^2} \approx \left|\frac{\Omega_{j}}{\delta_j}\right|^2
\label{eqRabiLeakage}
\end{equation}
where the first approximation is due to the use of perturbation theory and the second approximation is valid when $\Omega \ll \delta_j$.  Identical expressions are found for $|\braket{j|\hat{U}(t)|b}|^2$. Equation~\ref{eqRabiLeakage} describes the leakage out of states $\ket{1}$ and $\ket{e}$ during the $\ket{1}\leftrightarrow\ket{e}$ $\pi$-pulse, states $\ket{0}$ and $\ket{e}$ during the $\ket{0}\leftrightarrow\ket{e}$ $\pi$-pulse, and states $\ket{0;e}$ and $\ket{e;0}$ during the $\ket{0;e}\leftrightarrow\ket{e;0}$ exchange.  To treat leakage from $\ket{0}$ during the $\ket{1}\leftrightarrow\ket{e}$ $\pi$-pulse, let $\Omega=0$. Then \ref{eqLeakageGeneral} simplifies to
\begin{equation}
\left|\braket{j|\hat{U}(t)|0}\right|^2 \approx 2\left|\frac{\Omega_{j}}{\delta_j}\right|^2 (1-\cos{(\delta_j t)}).
\label{eqOtherLeakage}
\end{equation}
A careful choice of $\pi$-time reduces this leakage term, equivalent to minimizing leakage via alignment of power-spectrum zeros.

To evaluate the fidelity of each interaction step, we calculate the total population leakage probability from each nominally-populated state $\ket{i}$ as
\begin{equation}
p_i = \sum_j \left|\braket{j|\hat{U}(t)|i}\right|^2
\end{equation}
using Eq.~\ref{eqRabiLeakage} or \ref{eqOtherLeakage}, as appropriate.  The minimum fidelity is then
\begin{equation}
F \approx 1-\frac{1}{2}\max_i p_i.
\end{equation}

\end{document}